\documentclass[runningheads]{llncs}

\usepackage[T1]{fontenc}
\usepackage{graphicx}
\usepackage{subcaption}
\usepackage{wrapfig}

\usepackage[sorting=anyt,%
				giveninits=true,%
				maxbibnames=99,%
				citestyle=lncs,%
				natbib=true,%
				backend=biber,%,
				mincrossrefs=1000]{biblatex}

\AtEveryBibitem{
	\clearfield{editor}
	\clearfield{pages}
	\ifentrytype{book}{
	}{
	\ifentrytype{inbook}{}{
		\clearlist{publisher}
		\clearname{editor}
		\clearfield{isbn}
		\clearfield{issn}
	}}
	\ifentrytype{inproceedings}{
		\clearfield{series}
		\clearfield{volume}
		\clearfield{edition}
		\clearfield{doi}
		\clearfield{isbn}
		\clearfield{url}
		\clearfield{pages}
		\clearfield{month}
	}{}
	\clearlist{location}
	\clearlist{address}

}

\newcommand{\lastdate}[0]{May 2022\xspace}

\usepackage{listings}

\usepackage{wasysym}

\usepackage{dblfloatfix}
\usepackage[bottom]{footmisc}
\usepackage{longtable}
\usepackage{supertabular,booktabs}
\usepackage{lscape}
\usepackage{rotating}
\usepackage{calc}
\usepackage{balance}
\usepackage{lipsum}
\usepackage{tabularx}
\usepackage{xspace}
\usepackage[plain]{fancyref}
\usepackage[inline]{enumitem}
\usepackage{tabularx}
\usepackage{tabulary}
\usepackage{color}
\usepackage[prologue,table]{xcolor}
\usepackage{relsize}
\usepackage{threeparttable}
\usepackage{multirow}

\newcommand{\us}[0]{U.S.\xspace}
\newcommand{\uk}[0]{U.K.\xspace}

\begin{document}

\title{Back-to-the-Future Whois: An IP Address Attribution Service for Working with Historic Datasets}

\author{Florian Streibelt\inst{1} \and
Martina Lindorfer\inst{2} \and
Seda G\"urses\inst{3} \and
Carlos H. Ga\~n\'an\inst{3} \and
Tobias Fiebig\inst{1}}

\titlerunning{A Historic IP Attribution Service for Network Measurement}
\authorrunning{Streibelt et al.}

\institute{Max Planck Institute for Informatics
\email{\{fstreibelt,tfiebig\}@mpi-inf.mpg.de} \and
TU Wien
\email{martina.lindorfer@tuwien.ac.at} \and
TU Delft
\email{\{f.s.gurses,c.hernandezganan\}@tudelft.nl} }

\maketitle

\begin{abstract}
Researchers and practitioners often face the issue of having to attribute an IP address to an organization.
For \emph{current} data this is comparably easy, using services like whois or other databases.
Similarly, for historic data, several entities like the RIPE NCC provide websites that provide access to historic records.
For large-scale network measurement work, though, researchers often have to attribute millions of addresses.
For \emph{current} data, Team Cymru provides a bulk whois service which allows bulk address attribution.
However, at the time of writing, there is no service available that allows \emph{historic} bulk attribution of IP addresses.
Hence, in this paper, we introduce and evaluate our `Back-to-the-Future whois' service, allowing historic bulk attribution of IP addresses on a daily granularity based on CAIDA Routeviews aggregates.
We provide this service to the community for free, and also share our implementation so researchers can run instances themselves.
\end{abstract}

\section{Introduction}
\label{sec:introduction}

A common issue in the network measurement domain--but also in industry fields from Threat Intelligence to traffic engineering--is attributing an IPv4 or IPv6 address to an organization.
While, technically, Regional-Internet-Registries (RIRs) allocate IP addresses to organizations~\cite{RFC7020}, and provide a whois~\cite{RFC3912} infrastructure to make this information accessible, common whois interfaces are impractical for bulk requests.
This is mostly due to whois providing unstructured text data, which has to be appropriately parsed~\cite{RFC7485}.
Furthermore, organizations may have multiple organizational objects with overlapping and semantically equivalent data, which is not bit-equivalent or hides relationships due to subsidiaries from, e.g., different countries~\cite{caidaas2org}.
To address the needs of, especially, the threat hunting community, Team Cymru operates a bulk whois service, which allows users to bulk-request AS attribution for thousands of requests.

However, when working with \emph{historic} data-sets, sometimes ranging back decades, \emph{current} whois information may be ill suited to correctly attribute IP addresses, especially in the wake of IPv4 exhaustion~\cite{richter2015primer} and the accelerating IPv4 market~\cite{livadariu2013first,giotsas2020first,prehn2020wells,livadariu2017ipv4}.
Hence, in this paper, we introduce our historic whois service--Back-to-the-Future whois--which we implemented to address these challenges, leveraging the public CAIDA Routeviews aggregates~\cite{rv,caidaprefix2as}.
Our service is publicly available to the community at bttf-whois.as59645.net port tcp/43.
The service provides a historic address attribution service starting in May 2005 and for IPv4 and in January 2007 for IPv6.
It can be queried using a simple syntax, and provides structured JSON output, see also the website at \url{https://bttf-whois.as59645.net}.

\noindent In summary, we make the following contributions in this paper:
\vspace{-0.5em}
\begin{itemize}
	\item We introduce `Back-to-the-Future whois' (BTTF whois) as a public service for the research community as a simple way to historic attribute IPs. 
	\item We document our methodology, so researchers can independently distil historic IP attribution from Routeviews or the CAIDA aggregates.
	\item We evaluate BTTF whois' coverage over time on a case-study, and find BTTF to perform comparably to Team Cymru's bulk whois service on recent data, while outperforming it in accuracy for historic data.
\end{itemize}
\vspace{-0.5em}
\textbf{Structure:}
First, we introduce the datasets we use and our methodology for BTTF whois in \Fref{sec:dataset}. Next, we evaluate BTTF whois against Team Cumry's bulk whois in a sample case.
Finally, we first discuss our results and limitations in \Fref{sec:discussion}, before concluding in \Fref{sec:conclusion}.

\section{Dataset and Methodology}
\label{sec:dataset}
\vspace{-0.5em}

\subsection{Utilized Data}
\label{sec:whoisdata}

\paragraph{CAIDA Data for BTTF Whois}
The historic whois service leverages the aggregates of the RouteViews project compiled daily by CAIDA~\cite{caidaprefix2as}.
The dataset spans the time from May 2005 for IPv4 until today, and the time from January 2007 until today for IPv6, both with a daily resolution.
We use aggregates computed by CAIDA instead of aggregating the routing tables provided by the RouteViews project~\cite{rv} ourselves, as the RouteViews dataset is large (tenth of TB), and aggregation of this data is already a significant task in itself.

This prefix data alone is, however, insufficient to estimate a whois service based on routing data.
Routing data only maps IP addresses to ASes that announced the prefix at a specific time.
However, over time, ASes may change the organization they are allocated to.
Furthermore, we may find ASes that announce prefixes which are not registered to the announcing AS' organization, see for example Cogent announcing various customer prefixes,\footnote{\url{https://bgp.tools/as/174#prefixes}} see also \Fref{sec:lim}.

We address the issue of tying ASes to organizations by leveraging the AS2ORG dataset, also published by CAIDA~\cite{giotsas2014inferring,caidaas2org}.
The AS2ORG dataset covers the period from April 2004 up until today, with a quarterly resolution.
However, this reduced resolution will lead to a reduced reliability of the AS2ORG mappings, meaning that changes of ownership/authority over an AS may be reflected up to three months too late, while temporary changes of a duration less than three months may remain completely unnoticed, see \Fref{sec:lim}.

\paragraph{Case-Study Research Data}
To evaluate BTTF whois, we have to compare its efficacy against a `current' whois extract on a historic IP address dataset, where we can also investigate the impact of BTTF whois on the analysis results.
For this purpose, we use a study by Fiebig et al.\@ on the cloudification of universities~\cite{cloudheads}.
In their study they utilize the Farsight SIE DNS Dataset~\cite{farsight} -- specifically the \texttt{A}, \texttt{AAAA}, and \texttt{CNAME} records in the dataset -- from January 2015 to October 2022, to identify where universities' services are hosted.
The Farsight SIE dataset provides a \emph{historic} perspective on the IP addresses, names, and services under universities' domains.
For example: Finding \texttt{www.example.com. IN A 198.51.100.23} from January 2015 to April 2021 would indicate that the services was hosted in \texttt{TEST-NET-2} then. 
Finding only \texttt{www.example.com. IN A 203.0.113.11} from April 2021 onwards indicates that the service moved to \texttt{TEST-NET-3}.

To actually attribute IP addresses found via these records to ASes, Fiebig et al.\@ used -- in earlier iterations of the paper~\cite{cloudheadsold} -- the Team Cymru bulk whois service~\cite{teamcymru}.
Using that information, they then calculate the share of universities for several countries who have at least one system under their domain colocated with one of the big three cloud providers (Amazon, Google, Microsoft, or a combination of the three), see also \Fref{fig:apd_cymru}.
Overall, from January 2015 to October 2022, the dataset used by Fiebig et al.\@ spans a total of 880M DNS requests, pointing to between 500k and 6M individual IP addresses per month, adding up to 155M IPs (14M unique). Naturally, the same IP address may occur in several months.
As Fiebig et al.\@ initially used the Team Cymru bulk whois service (only containing \emph{current} AS attributions), their work is an ideal case study to evaluate how using a historic AS attribution service influences observed results.

\paragraph{Team Cymru Whois Data}

As a base-line, we requested bulk whois data from Team Cymru's bulk whois service for all unique addresses in January 2023.
We used the Team Cymru whois to resolve all 14M unique IP addresses in the university dataset.
For each IP address the bulk whois service of Team Cumry returns the currently associated AS number, the requested address, and the AS Name and location of the corresponding AS.

\subsection{Methodology}
\label{sec:whoismethod}
In this section, we describe how we organized the CAIDA AS2ORG and AS2Prefix datasets in our service daemon to enable quick queries for individual addresses against the dataset.
The major challenge--preventing a traditional RDBMS from being used--is that these datasets contain whole prefixes, instead of individual IP addresses, and relations between objects are complex.
This would lead to, for example in SQL, a nested JOIN structure which limits performance of an RDBMS.
To prevent this bottleneck, our implementation uses a completely in-memory prefix trie, i.e., pytricia~\cite{pytricia}.

\paragraph{AS2ORG Data-Structure.}
To use the supplied dataset to identify the AS and organization announcing a specific IP address, we first create a data-structure mapping time-frames, organizations, and ASes to each other.
The challenge here is that the resolution of the supplied data is relatively low.
Furthermore, we find that the supplied data regularly contains parsing errors, as it has been sourced from RIR supplied whois data, which is known to be often unstructured and to have volatile formats~\cite{liu2015learning}.

To handle the sparseness of the supplied data, we do have to make decisions on the margin of error that is acceptable for a whois service when making an educated guess for the organizational affiliation of an AS in between two quarterly files.
There, we have to handle four cases:
\begin{itemize}[leftmargin=*]
\item \textbf{AS2ORG unchanged:} If, in both files, the AS is mapped to the same organization, we assume that it was continuously mapped to the same organization between the two dates for which we have data.
\item \textbf{AS missing from newer file (AS removed):} If an AS has been removed, we consider it to be removed from the day directly following the last quarterly file's date in which the AS could be found.
\item \textbf{AS missing from older file (AS added):} If an AS has been added, we consider this AS mapping to be valid from the date of the file in which the AS first occurs (again).
\item \textbf{AS2ORG changed:} If the AS2ORG mapping changes between two adjacent files, we consider this change to have come into effect on the day after the older files' collection date.
\end{itemize}
Following this approach, we can then construct a continuous mapping of ASes to organizations in our data-structure.

\paragraph{Prefix Tree (Trie).}
Next, we iterate through the list of available files by date, and add the prefixes we find to an IP trie~\cite{pytricia}.
In that trie, each added prefix holds a list at date ranges when it was observed.
For each prefix in our input files, we check if the prefix exists in the trie.
Here, we have to handle four cases:
\begin{itemize}[leftmargin=*]
\item \textbf{Prefix is not in the trie:} We add the prefix to the trie, setting the 'first seen' field to the date of the collection date of the currently processing file.
\item \textbf{Prefix is in the trie:}
\begin{itemize}
\item \textbf{No gap to last-seen date:} If the last-seen date of the prefix is the date of the day before the collection time of the currently processing file, we update the last-seen date of the most recent date-range to the date of the currently processing file.
\item \textbf{Gap to last-seen date:} If the last-seen date of the prefix is not the date of the day before the collection time of the currently processing file, we add a new date-range to the list of date-ranges, and set the first seen date to the date of the currently processing file.
\item \textbf{Originating AS changed:} If the originating AS(es; see below) changed from the last seen state, we treat the prefix as a new prefix, i.e., start a new date range associated with the new ASes.
\end{itemize}
\end{itemize}
In all cases, the prefix is attributed to the ASes we observe as announcing the prefix.
There, we also have to handle several special cases:
\begin{itemize}[leftmargin=*]
\item \textbf{Prefix originated by exactly one AS:} If a prefix is originated by exactly one AS, we add this AS as the authoritative AS.
\item \textbf{MOAS prefix:} If a prefix is announced by multiple ASes at the same time, commonly known as a MOAS (Multi Origin AS) prefix, we add all these ASes to the announcement state, see the section on handling requests for details on the presentation.
\item \textbf{ASSET aggregate:} ASes may aggregate prefixes received from downstream ASes. Fore example, if AS65536 announces 198.51.100.0/25 to AS65538, and AS65537 announces 198.51.100.128/25 to AS65538, AS65538 can aggregate these announcements to 198.51.100.0/24, only announcing that to its peers, while also aggregating AS65536 and AS65537 to \{ AS65536, AS65537 \} in the AS path of that announcement. The information whether 198.51.100.0/25 was originated by AS65536 or AS65537 is lost in this process. As this is suggested to occur only on provider aggregatable IP space~\cite{RFC2519}, we attribute the whole /24 to the aggregating AS, i.e., AS65538 in this case.
\end{itemize}
After having determined the ASes to which we attribute a prefix, we look up the associated AS2ORG mapping from our first datastructure and add that information to the date range.
Please note that the trie data structure handles the occurrence of more specific prefixes by a branching approach, i.e., we can add 198.51.100.128/25 to the trie, even if 198.51.100.0/24 is already present.
When looking up addresses, the more specific will match, and we will have to traverse the tree upward, see also below under `Lookups'.
Loading the full data set into the implementation takes around 24 hours.

\paragraph{Filtering.}
Prefix announcements on the Internet are noisy.
Specifically, we may regularly observe organizations announcing prefixes they are not supposed to announce~\cite{sermpezis2018survey}, announce prefixes that are more specific than the maximum agreed prefix size in the global routing table (/24 for IPv4 and /48 for IPv6)~\cite{sediqi2022hyper}, announce prefixes that are unreasonably short, e.g., when leaking default routes, or announce prefixes and AS numbers from reserved ranges~\cite{RFC1918} (see also IANA's registires\footnote{\url{https://www.iana.org/assignments/iana-ipv4-special-registry/iana-ipv4-special-registry.xhtml},}\textsuperscript{,}\footnote{\url{https://www.iana.org/assignments/iana-ipv6-special-registry/iana-ipv6-special-registry.xhtml}}).
Reserved prefixes are statically added to our lookup daemon, and reported as such upon lookup.
Hence, when importing prefixes we are filtering all announcements less specific than a /8 for IPv6 and /18 for IPv6, and more specific than a /24 for IPv4 and /48 for IPv6.
Similarly, we exclude all prefixes originated by private and reserved AS numbers, i.e., 0~\cite{RFC6483, RFC7607}, 23456~\cite{RFC6793}, 64496-64511~\cite{RFC5398}, 64512-65534~\cite{RFC1930,RFC6996}, 65535~\cite{RFC7300}, 65536-65551~\cite{RFC5398,RFC6793}, 65552-131071 (IANA Reserved), 4200000000-4294967294~\cite{RFC6996}, and 4294967295~\cite{RFC7300}.

\paragraph{Lookups.}
The implementation of the historic whois service allows lookups with daily granularity.
When an IP address or prefix is looked up, we first identify the most specific match.
Next, we check if the prefix has been announced at the given date, i.e., if it has a date-range covering the requested date.
If it does not have a corresponding date range, we traverse the tree until we either find a less specific prefix with a covering date-range or arrive at the root of the address tree.
If we reach the root, we return that the prefix was not found at that date.

For the most specific prefix with a covering date range, we return the requested IP address or prefix, the requested date, and the result set. 
The result set contains the dates when the prefix was first and last observed for the date-range covering the requested date, with the last-seen date being null if the prefix was still being observed in the newest file imported into the daemon.
Additionally we return the identified prefix and the list of ASes associated with the prefix.
For each AS we also return an AS2ORG mapping, listing the ASN, the ASNAME and RIR where the ASN has been registered.
Furthermore, we return all organizations associated with the AS at the time of the request, which includes the country code registered for the organization, the RIR the organization object has been obtained from, and the name of the organization.

\paragraph{Implementation, Infrastructure, and Performance.}
We implemented the historic whois system in a team using roughly three person months between May and August 2022 in Python.
To handle our request load, we deployed forty instances behind a load-balancing frontend on a cluster of four hardware machines. 
Each instance consumes roughly 16GB of memory (including caches) and has access to two dedicated CPU threads, leading to a total resource consumption of 80 CPU cores and 640GB of memory, without Kernel Same-Page Merging (KSM) applied.
An instance can process around 1.2K lookups a second, allowing us to perform the address resolution for the 155M addresses over 7 years in a bit more than 1.5 hours given noise in actual lookup rates and a maximum parallelization factor of 40.

\section{Results}
\label{sec:results}
In this section, we describe how we evaluate the efficacy of BTTF whois using the work of Fiebig et al.~\cite{cloudheads} as a case-study.
We first introduce the results Fiebig et al.\@ obtained by using Team Cymru's bulk whois service.
We then compare the attribution of address ownership between Team Cymru's bulk whois service and BTTF whois.
Finally, we revisit the results of Fiebig et al.\@, and describe how using BTTF whois influences them.

\subsection{Universities' Cloud Usage: Team Cymru's Bulk Whois}
\begin{figure}[t!]
        \begin{center}
                \includegraphics[width=.91245\columnwidth]{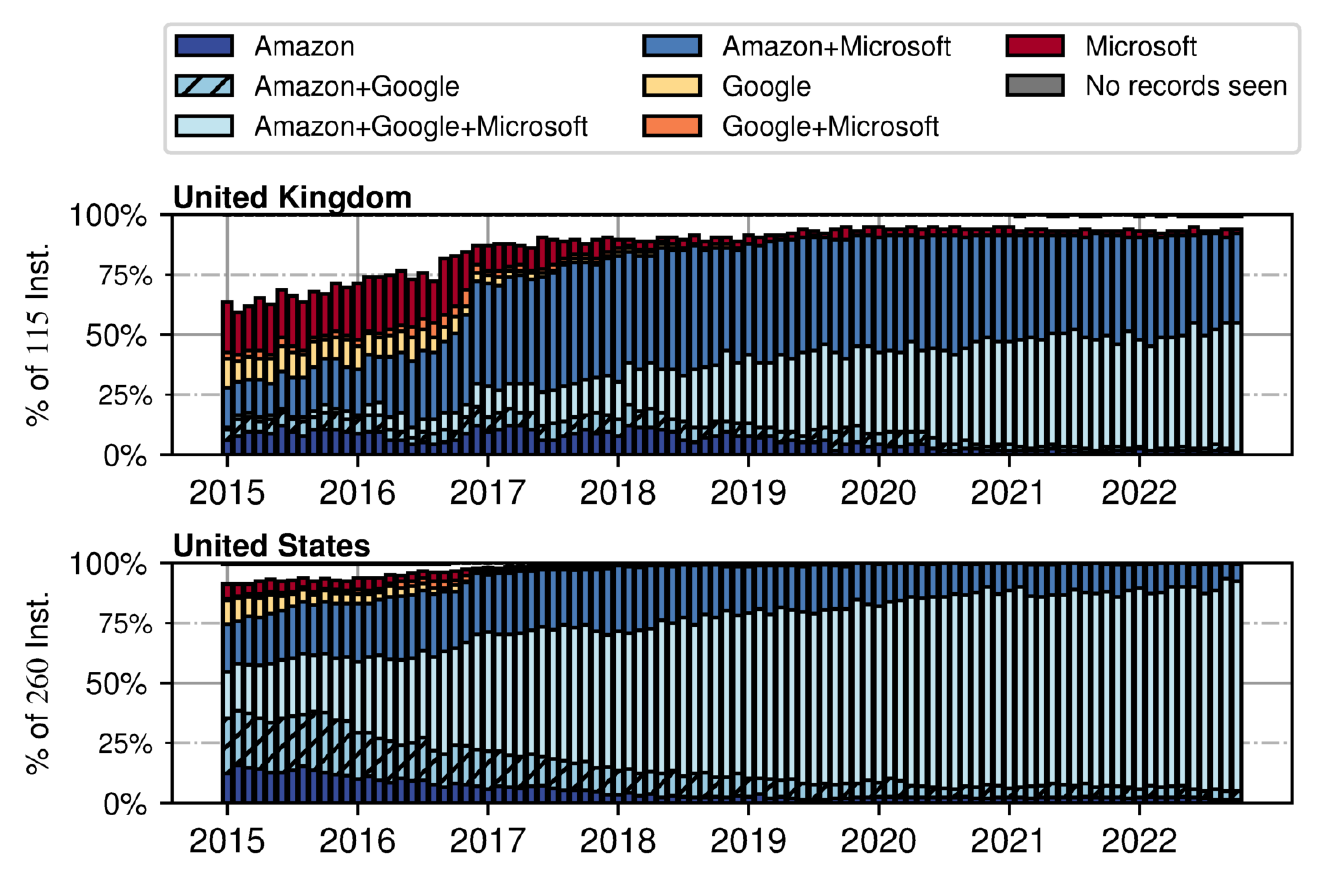}
				\vspace{-.5em}
                \caption{Cloud use attribution for universities in the \uk and the \us (January 2015--\lastdate) based on Team Cymru bulk-whois data.}
                \label{fig:apd_cymru}
        \end{center}
		\vspace{-1em}
\end{figure}

As outlined in \Fref{sec:whoisdata}, Fiebig et al.\@ use the Farsight SIE dataset to identify IP addresses to which names under universities domains point with a monthly granularity.
Using Team Cymru's bulk-whois service, they then attribute these IPs to AS numbers.
For their final analysis, they then calculate the share of universities in a country under whose domains at least one name ultimately points to an IP address announced by one of Amazon's, Google's, or Microsoft's ASes.
Naturally, a university may have multiple names under its domain that point to addresses announced by different cloud providers.
\Fref{fig:apd_cymru} depicts their results from January 2015 to October 2022 for 115 \uk universities and 260 \us universities, with each bar in the bar-plots representing the distribution observed during a single month.

For both, the \us and the \uk, they find an overall high prevalence of at least one service or site being run on Amazon, Google, or Microsoft systems.
Notably, the \us already shows an over 90\% saturation in cloud use, with the main development being that the prevalence of universities having infrastructure located at all three major cloud providers continuously rises over time.
For the \uk, still around 75\% of universities have at least one service in the major three clouds in January 2015, followed by a gradual increase across all platforms.
Still, even in October 2022, the use of Google systems in the \uk is lower comparison to the observed \us usage.

\subsection{IP Attribution Comparison: BTTF Whois vs. Team Cymru}
\begin{figure}[t!]
        \begin{center}
                \includegraphics[width=\columnwidth]{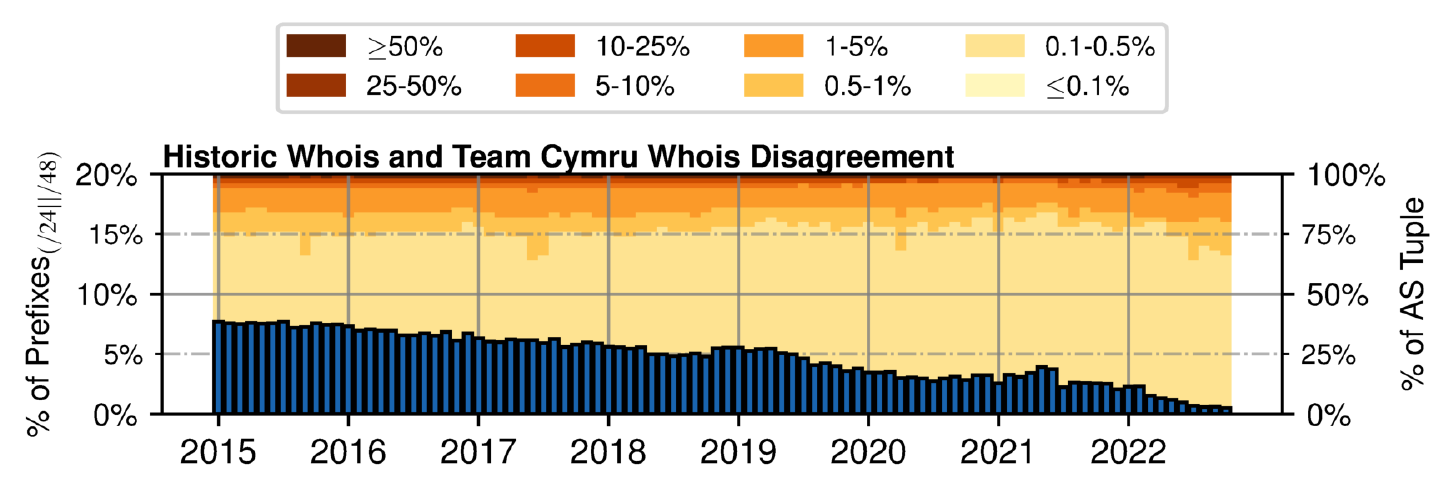}
				\vspace{-.5em}
                \caption{Percentage of prefixes in the dataset on which the historic whois service we implemented and the data from Team Cymru's bulk whois service disagree.
				The shaded background indicates the distribution of disagreement over AS tuple, i.e., the tuple of the ASes to which Team Cymru attributes a prefix and the ASes BTTF attributes a prefix to.
				Note that, as common with centralization, only a minor fraction of AS tuples is responsible for the bulk of disagreement.}
                \label{fig:comp}
        \end{center}
		\vspace{-1em}
\end{figure}

We first compare the direct attribution results between Team Cymru's bulk whois and BTTF whois.
For that, we first use Team Cymru's bulk whois to attribute all 14M unique IP addresses found by Fiebig et al.\@ (see \Fref{sec:whoisdata}) to ASes.
We then use the BTTF whois service to attribute all addresses seen in a month to ASes based on the joined state seen on the 1\(^{st}\), 14\(^{th}\), and 28\(^{th}\) of that month.
Finally, we calculate the disagreement in attribution between the two data sources over time per /24 (IPv4) and /48 (IPv6), the minimum prefix sizes that can be successfully announced.
We strictly compare the sets of ASes, only considering an exact match to be agreement.
If one service returns a subset of ASes of the other, we consider this a disagreement.

The intuitive assumption for this is that Team Cymru's whois data is accurate \emph{`as of now'}, while accuracy declines gradually while going further into the past as prefixes have been transferred between organizations.
Based on these assumptions, there should be a high agreement between data from the historic whois service and the Team Cymru provided data for relatively recent months.
However, the discrepancy should increase when we go further back in time.

As \Fref{fig:comp} depicts, this is indeed the result we obtain.
While disagreement started out at around 7.67\% in 2015, it continuously decreases over time, with the lowest disagreement occurring in October 2022, with 0.49\%.
Overall, this result aligns with our predictions in terms of reliability for the historic whois service.
Hence, as it comparably reliable on data where Team Cymru's whois is reliable, we assume BTTF whois to be reliable for historic data as well.

\subsection{Impact of BTTF Whois on Case-Study Analysis}

\begin{figure}[t!]
        \begin{center}
                \includegraphics[width=.91245\columnwidth]{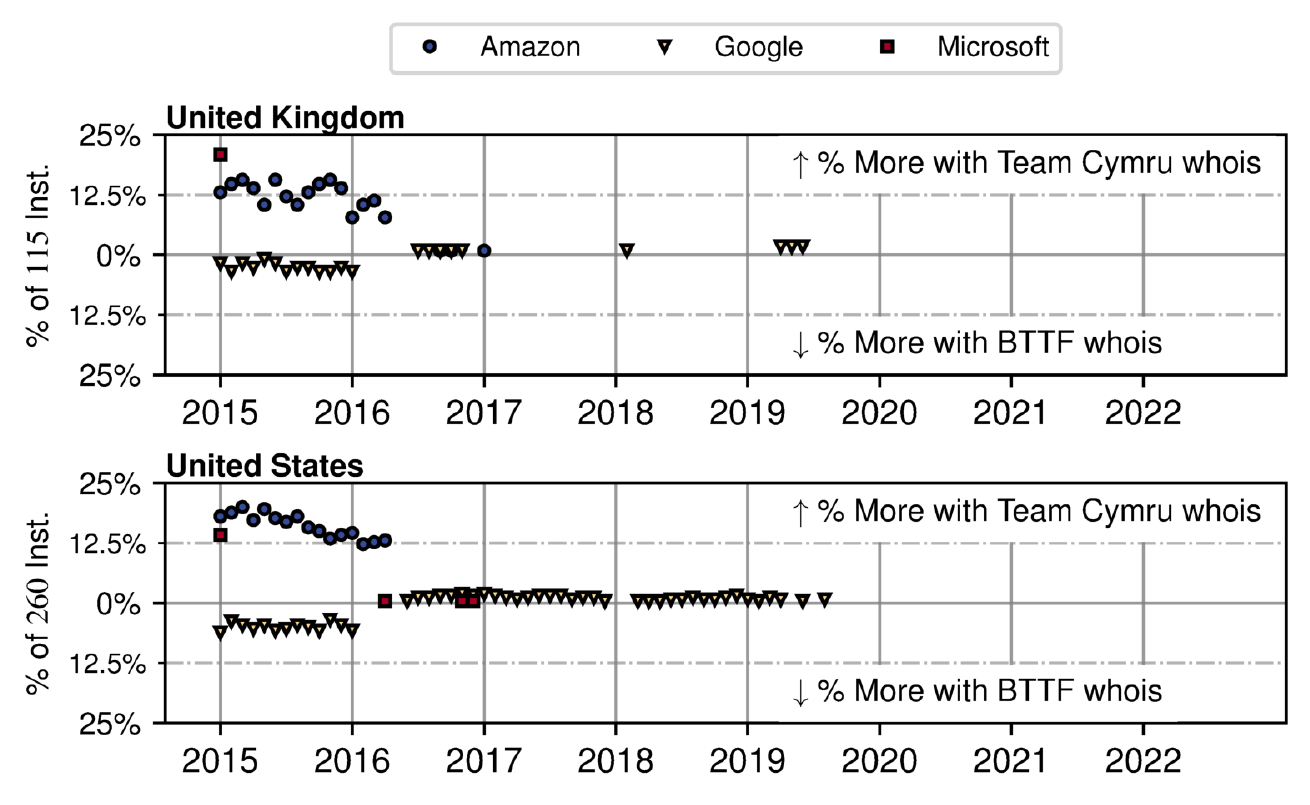}
				\vspace{-.5em}
                \caption{Difference in cloud use attribution for universities in the \uk and the \us (January 2015--\lastdate) between Team Cymru bulk-whois data and historic bulk-whois data as absolute percent values as relative change considering Team Cymru as the base-line, i.e., positive values mean \emph{more based on Team Cymru whois data}, while negative values mean \emph{more based on historic bulk whois data}.}
                \label{fig:apd_diff}
        \end{center}

		\vspace{-1em}
\end{figure}
For demonstrating the benefits of our historic whois service, we analyze how its different perspective on IP address ownership influences the results Fiebig et al.\@ presented~\cite{cloudheads}.
To this end, we compared the final cloud hosting verdict for several countries between an analysis where our historic whois service has been used and one where Team Cymru's whois has been used (see \Fref{fig:apd_diff}).
Over all countries in our analysis, we only observe a significant impact in the \uk and the \us.
For the \uk and the \us, we find that, overall, the number of universities attributed to Amazon (i.e., Amazon, Amazon+Google, Amazon+Microsoft, Amazon+Google+Microsoft) are estimated higher by data from Team Cymru's whois until May 2016 by around 12.5\%.
Additionally, we find a minor (\(\leq\)5\%) underestimation for Google use in 2015, and a high overestimation of Microsoft use in January 2015 only.

Focusing on the Amazon case, we were able to attribute it to 18.0.0.0/8, the IPv4 address block formerly allocated to the Massachusetts Institute of Technology (MIT). 
In 2017, MIT announced its intent to sell large parts (87.5\%) of this address block to Amazon~\cite{MITref}. 
The transfer of addresses was finalized in 2019, with the creation of associated route objects~\cite{amzroute}, but the networks to be sold were cleared ahead of time.
As several Universities in the \us and \uk had names under their domain pointing to IP addresses from MIT, and -- based on \emph{currently} accurate attribution information -- these now belong to Amazon, these addresses were wrongly attributed to Amazon.

\begin{figure}[t!]
        \begin{center}
                \includegraphics[width=.91245\columnwidth]{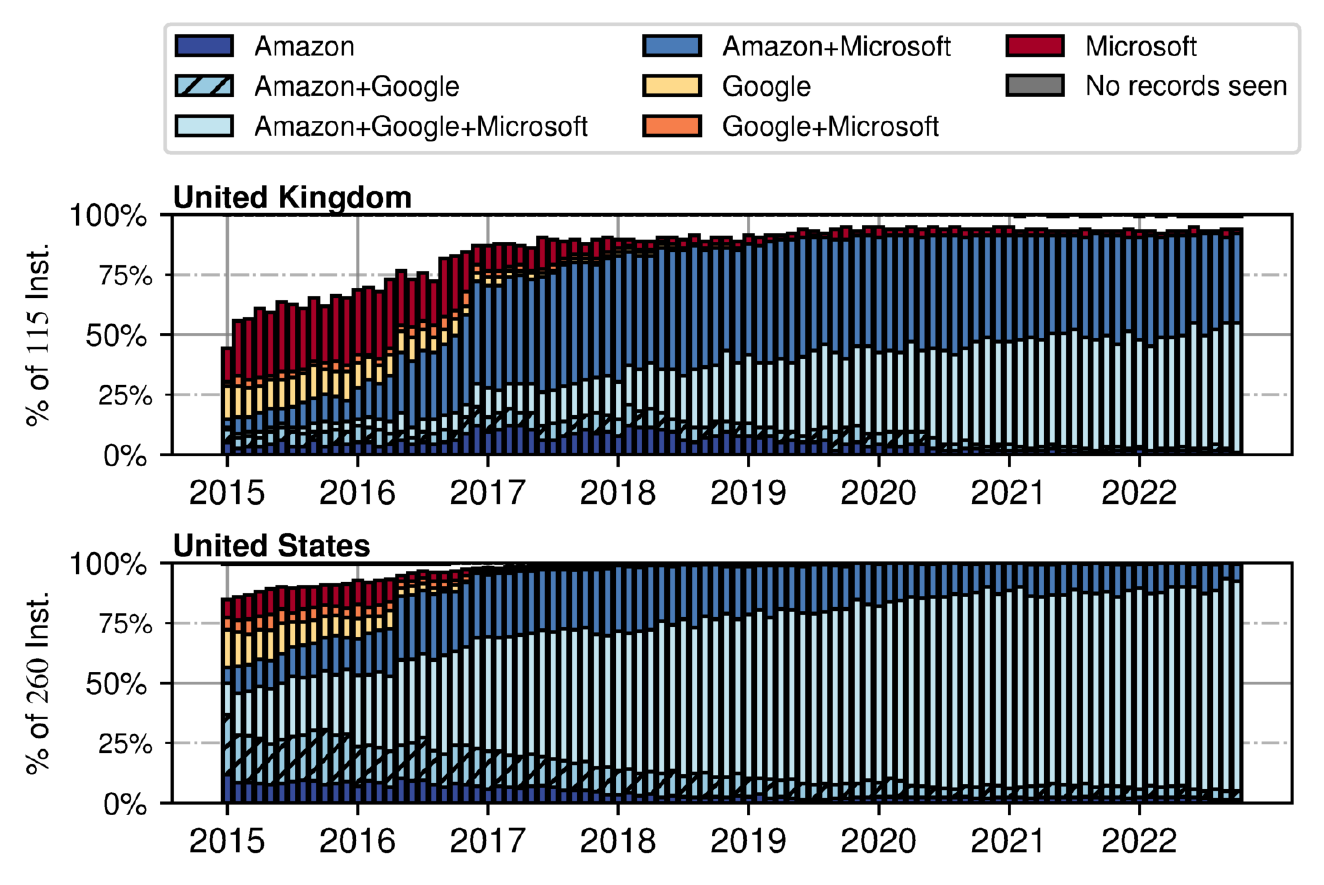}
				\vspace{-.5em}
                \caption{Cloud use attribution for universities in the \uk and the \us (January 2015--\lastdate) based on historic bulk-whois data.}
                \label{fig:apd_bttf}
        \end{center}
		\vspace{-2em}
\end{figure}

To better understand the significance of this attribution error, we compare the cloud usage graphs generated when using whois data sourced via the Team Cymru whois service (see \Fref{fig:apd_cymru}) with the updated version relying on our historic whois (see \Fref{fig:apd_bttf}).
We find that for both countries, the \us and the \uk, using the historic whois service reveals an initially lower usage of Amazon based hosting, followed by a more rapid increase.
For example, in the \us, we find that the initial share of universities also using Amazon hosted services now hovers around 60\% instead of the 75\% initially observed.

In the \uk the effect has been more pronounced.
Instead of the gradual increase initially assumed based on Team Cymru's whois data, we now a lower share of Amazon service usage for the \uk in 2015 (around 12.5\% instead of 25\%), increasing during 2016.
%This quickly increased over the course of, especially, 2016.
Hence, by initially not using a historic whois service, Fiebig et al.\@ missed an important growth effect in the data.

\section{Discussion}
\label{sec:discussion}
In this section, we discuss lessons learned for research on historic datasets, discuss the limitations of our approach, and outline further work.

\subsection{Lessons Learned for Research on Historic Datasets}
In \Fref{sec:results}, we have seen the major impact incorrect address attribution can have on research results when working with historic records.
Especially research that investigates research questions in which IP address ownership and control is instrumental--as the case-study work by Fiebig et al.--becomes more robust by selecting a more accurate IP address attribution methodology.
Given the growing availability of historic datasets containing IP addresses, for example, the Farsight SIE dataset~\cite{farsight}, the OpenINTEL dataset~\cite{hohlfeld2018operating,van2016high}, but also historic trace-route datasets~\cite{luckie2008traceroute}, or IXP datasets~\cite{chatzis2013benefits}, we expect more future research to deal with historic address datasets.
At the same time, the exhaustion of IPv4~\cite{richter2015primer}, and the associated growth of the IP address and leasing market~\cite{livadariu2013first,giotsas2020first,livadariu2017ipv4} will make real-time whois information increasingly unreliable for such historic datasets.
As such, our service fills an important gap for the research community.

\subsection{Limitations}
\label{sec:lim}
Despite our successful validation of the historic whois service by demonstrating it performs comparable to established bulk whois services on recent data, there are several limitations which should be discussed.
First, the utilized CAIDA data exhibits several inconsistencies in data, e.g., AS numbers having a dot in the middle.
The CAIDA prefix data is an aggregate of RouteViews data.
The aggregation process may occlude specific announcements, e.g., if a prefix is not yet visible at the single route collector used by CAIDA.
Similarly, prefix hijacks~\cite{sermpezis2018survey} may inject routes into the aggregate table, which are then wrongly attributed to the hijacking organization.
This issue could only be addressed by a more elaborate data structure, that includes Internet Registry Routing object data as well as RPKI~\cite{RFC6483} data -- which is difficult to obtain in historic form -- and heuristics to identify and exclude route hijacks.
Given the current accuracy of the historic whois service, we consider this approach as out-of-scope.

Furthermore, there are several limitations in the AS2ORG mapping data.
AS family calculation~\cite{caidaas2org}, i.e., grouping of ASes to a common organization if the organizational objects of theses ASes are, e.g., subsidiaries of a common corporation, are unreliable over time.
Fields from the whois data provided by RIRs is not consistently parsed, and fields contain faulty data if the base format on the RIRs side changes without the parsers that generate the AS2ORG extract we rely on being adjusted.
In addition, the AS2ORG maps have a quarterly granularity, which makes AS2ORG attribution unreliable when changes occur, as discussed in our methodology.

Finally, as noted before, a prefix being announce by an AS does not necessarily mean that this prefix is allocated to, or owned by said AS, see the example of announcements of AS174.
Hence, our implemented historic whois service will mis-attribute prefixes that are registered to an organization that is not the organization to which the announcing AS is associated.

Nevertheless, again given the observed reliability in comparison with Team Cymru's whois service, we consider the current implementation of our historic whois service as sufficiently robust to provide historic whois data.
Effectively, it is comparably accurate to the commonly used Team Cymru  whois service on recent data, while providing higher accuracy in historic data, as highlighted by the case of MIT's /8 network.

\subsection{Future Work}
As discussed in our limitations section, our reliance on the CAIDA aggregates of the Routeviews BGP announcement collections still limits the accuracy of our data.
To improve our service, it would hence be advisable to not only provide routing information based IP attribution, but also access other sources for historic whois information, and attach it to returned records if it is available.
For example, RIPE NCC provides a historic non-bulk whois service.
We are in conversations with relevant RIRs and registrars to obtain access to these datasets, so that our service can--along with routing based attribution information, i.e., the announcing AS--also return information from RIR databases.
If these datasets become available, it would also be prudent to compare RIR information with actual routing information over the historic timeframe covered by our service.

Similarly, Routeviews data is available for a longer timeframe than the CAIDA aggregates. 
Hence, we also plan to aggregate Routeviews information from before the first CAIDA aggregates became available--as early as 2000--to include in our BTTF whois service.

\section{Conclusion}
\label{sec:conclusion}

In this paper, we introduce and evaluate BTTF whois as a public community service.
This historic whois service allows more accurate estimations of IP address ownership, especially when the concerned IP address has been observed in the past.
Based on a case-study, we demonstrate how the use of an accurate historic whois service allows deeper insights into datasets, and reveal developments that would remain shrouded when only relying on \emph{current} whois information.

Nevertheless, several challenges exist, which should be resolved in further iterations of the development of our service.
This includes aggregating the RouteViews dataset ourselves -- especially as older data-sets are available than aggregated by CAIDA -- and continuously collecting RIR provided data for generating AS2ORG maps ourselves, including addressing the issue of organizational families more reliably.
Furthermore, future implementations should include IRR and RPKI data to make the implementation more robust against data noise due to prefix hijacks and the announcement of prefixes by ASes not belonging to the prefix-holder's organization.

\noindent
\textbf{Service Availability: } You can use a publicly available instance of BTTF whois at bttf-whois.as59645.net port tcp/43. See Appendix~A for usage details and \url{https://bttf-whois.as59645.net} for further information.

\vspace{1em}
\noindent\textbf{Acknowledgements:}
We thank Farsight Security, Inc. (now DomainTools) for providing access to the
Farsight Security Information Exchange's passive DNS data feed.  Without this
data, the project would not have been possible.  The authors express their
gratitude to the anonymous reviewers and our shepherd Thomas Krenc for their
thoughtful and encouraging input during the reviewing process.  Furthermore, we
thank the reviewers who accompanied \emph{`Heads in the Clouds? Measuring
Universities' Migration to Public Clouds: Implications for Privacy \& Academic
Freedom'} at PoPETS from 2022.2, via 2022.4, to 2023.2 for seeding the idea to
implement the BTTF whois service and their continuous encouragement to pursue
this work. Finally, Sebastian Lohff's input on the implementation and
performance tuning were invaluable to realize the service in a production-ready
manner.  This work was partially funded by the German Federal Ministry of
Education and Research under the project 6G-RIC, grant 16KISK027.  Any
opinions, findings, and conclusions or recommendations expressed in this
material are those of the authors and do not necessarily reflect the views of
Farsight Security, Inc., DomainTools, the German Federal Ministry of Education
and Research, or the authors' host institutions and affiliations.

{

\printbibliography
}

\appendix
\section{BTTF Whois Short Documentation}
Here, we document a) how you can use BTTF whois with a whois client, and b) how to obtain bulk results.
Furthermore, we provide an overview over the returned JSON's structure.

\subsection{Using BTTF Whois Manually}
BTTF whois can be used with a standard \texttt{whois} client.
The date format is YYYYMMDD.
%\begin{lstlisting}[float,floatplacement=T,caption=Manual use of BTTF whois.\vspace{-1em},label=lst:manual]
\begin{verbatim}
% whois -h bttf-whois.as59645.net '1.1.1.1 20210101'
# This is the historic IP to AS mapping service
# Contact: <contact@as59645.net>
# Trie Status: READY - loaded 2191224 IPv4 and 345010 IPv6 prefixes
# AS2Org Status: 119641 AS and 201610 organisations loaded
# Enter HELP to get basic usage information
# NOTICE: OUTPUT FORMAT: JSON-SHORT
# READY
{
    "ipaddr": "1.1.1.1",
    "qdate": "20210101",
    "results": {
        "timestamp": 20180320,
        "until": null,
        "prefix": "1.1.1.0/24",
        "aslist": [
            13335
        ],
        "orgmapping": {
            "13335": [
                {
                    "asn": 13335,
                    "aut": {
                        "aut": 13335,
                        "aut_name": "CLOUDFLARENET-AS",
                        "org_id": "@family-471",
                        "opaque_id": "",
                        "source": "RIPE"
                    },
                    "seen": [
                        "20180703"
                    ],
                    "changed": "20180703",
                    "change_guessed": true,
                    "orgs": [
                        {
                            "org_id": "@family-471",
                            "org": {
                                "org_id": "@family-471",
                                "org_name": "Cloudflare Inc",
                                "country": "US",
                                "source": "ARIN,RIPE"
                            },
                            "seen": [
                                "20180703"
                            ],
                            "changed": "20180703",
                            "change_guessed": true
                        }
                    ]
                }
            ]
        }
    }
}
\end{verbatim}
%\end{lstlisting}

\subsection{Using BTTF Whois for Bulk Requests}
BTTF whois ingests bulk requests enclosed in a `begin` and `end` statement:
\begin{verbatim}
% cat ./file
begin
1.1.1.1 20210101
1.1.1.1 20120101
8.8.8.8 20210201
end
\end{verbatim}
You can use \texttt{netcat}/\texttt{nc} to send this file to the bulk whois service and receive the results directly or redirect them to a file:
\begin{verbatim}
% cat ./file | nc bttf-whois.as59645.net 43
# This is the historic IP to AS mapping service
# Contact: <contact@as59645.net>
# Trie Status: READY - loaded 2169926 IPv4 and 319033 IPv6 prefixes
# AS2Org Status: 119005 AS and 199948 organisations loaded
# Enter HELP to get basic usage information
# NOTICE: OUTPUT FORMAT: JSON-SHORT
# READY
{"IP": "1.1.1.1", "QDATE": "20210101", "results": {"DATA_FIRST": [...]
{"IP": "1.1.1.1", "QDATE": "20120101", "results": []}
{"IP": "8.8.8.8", "QDATE": "20210201", "results": {"DATA_FIRST": [...]
# goodbye
\end{verbatim}

\subsection{BTTF Whois JSON Data Structure}
Below, you can find an overview of the response fields returned by BTTF whois.
\begin{verbatim}
{ # Requested IPv4 or IPv6 address
  "IP": "1.1.1.1",
  # Date for which data was requested
  "QDATE": "20210101",
  "results": {
    # First time the most specific prefix for address has been seen 
    # first with this specific set of announcing ASes
    "DATA_FIRST": 20180320,
    # Last time this entry was seen, i.e., valid until. If it is 
    # null, the most specific is still visible in the most recent
    # dataset (valid NOW).
    "DATA_LAST": null,
    # List of ASNs that announced the most specific prefix for the
    # requested address.
    "asns": [
      13335
    ],
    # The most specific matching prefix from the dataset.
    "prefix": "1.1.1.0/24",
    # AS2ORG mappings for all announcing ASN.
    "as2org": [
      {
        # AS number
        "ASN": 13335,
        # AS name
        "ASNAME": "CLOUDFLARENET-AS",
        # RIR that is the data source in the AS2ORG mappings
        "RIR": "RIPE",
        # Org objects associated with the ASN
        "orgs": [
          {
            # Country code attributed to an organization
            "CC": "US",
            # RIRs that hold an instance of this ORG object
            "RIR": "ARIN,RIPE",
            # Organization name from the ORG object
            "ASORG": "Cloudflare Inc"
          }
        ]
      }
    ]
  }
}
\end{verbatim}

\end{document}